\documentclass[manuscript,epsfig]{aastex}
\onecolumn

\begin{document}

\newcommand{\bedm}{\begin{displaymath}}
\newcommand{\eedm}{\end{displaymath}}
\newcommand{\be}{\begin{equation}}
\newcommand{\ee}{\end{equation}}
\newcommand{\etal}{{et~al.~}}

\slugcomment{Submitted to The Astrophysical Journal}

\shortauthors{Doroshkevich \etal}

\shorttitle{Ionization history}

\title{Ionization history of the cosmic plasma in the light of the 
recent CBI and future PLANCK data.}

\author{ A.~G.Doroshkevich\altaffilmark{1},
         I.~P.~Naselsky\altaffilmark{2},
         P.~D.~Naselsky\altaffilmark{1,2},
        I.~D.~Novikov\altaffilmark{1,3,4,5}}

\begin{abstract}
The paper is devoted to the methods of determination of the
cosmological parameters and distortions of ionization history 
from recent CMB observations.
We show that the more complex models of kinetics of recombination
with a few "missing" parameters describing the recombination process
provide better agreement between the measured and the expected 
characteristics
of the CMB anisotropy. In particular, we consider the external sources
of the resonance and ionizing radiation and the model with the strong 
clustering of the baryonic component. These factors can constrain
the estimates of the cosmological parameters usually discussed. We
demonstrate also that the measurements of the CMB polarization can 
improve these estimates and, for the precision expected from the 
PLANCK mission, allow to discriminate a wide class of models.
\end{abstract}
\keywords{theory:  --
cosmic microwave background -- ionization}

\altaffiltext{1}{Theoretical Astrophysics Center, Juliane Maries Vej
30, DK-2100,  Copenhagen, Denmark.}
\altaffiltext{2}{Rostov State University, Zorge 5, 344090 Rostov-Don, 
Russia.}
\altaffiltext{3}{Astro Space Center of Lebedev Physical Institute,
Profsoyuznaya
84/32, Moscow, Russia.}
\altaffiltext{4}{University Observatory, Juliane Maries Vej 30,
DK-2100,  Copenhagen, Denmark.}
\altaffiltext{5}{NORDITA, Blegdamsvej 17, DK-2100,  Copenhagen, Denmark.}
\maketitle
\onecolumn

\section{Introduction}
Recent observations of the CMB anisotropy such as the BOOMERANG
(de Bernardis et al. 2000), the  MAXIMA-1 (Hanany et al. 2000),
the  DASI (Halverson et al. 2001), the VSA (Watson et al. 2002), 
the Archeops (Benoit et al. 2002) and especially the new CBI data 
(Mason et al. 2002) and  the DASI polarization measurements (Kovac 
et al. 2002) provide the good base for the progress in understanding 
the most general properties of the early Universe. In these 
experiments, major attention is on the statistical
properties of the signal and the determination of the
power spectrum of the CMB anisotropy and polarization. The obtained 
cosmological parameters for adiabatic cold dark matter model (CDM) 
are well consistent with the Big Bang nucleosynthesis predictions, SN Ia 
observational data and results obtained from investigations of galaxy 
surveys. Further progress will be achieved with more  sensitive  
experiments such as the launched MAP and the upcoming PLANCK missions. 
They will be able to measure the CMB anisotropy and polarization
power spectra up to the multipole range $l\simeq 3-4\cdot 10^3$ where 
the lensing effect, Sunyaev-Zel'dovich effect in 
clusters, secondary anisotropies and some peculiarities of the 
ionization history of the Universe dominate. 

The high precision of the future CMB measurements allows to
reveal some distortions with respect to the standard model of the 
hydrogen recombination process occurred at redshifts $z\simeq 10^3$.
Here we show that polarization power spectrum contains important 
information about the kinetics of hydrogen recombination and allows 
to trace the ionization history of the cosmic plasma in wide range 
of redshifts  $10^3\geq z\geq 6-10$.

For the baryon--dominated Universe the classical theory of 
the hydrogen recombination was developed in Peebles (1968) 
and Zel'dovich, Kurt and Sunyaev (1968). For the dark matter 
dominated Universe, it was generalized in Zabotin and Naselsky 
(1982), Jones and Wyse (1985), Seager, Sasselov and Scott (1999) 
(see detailed review in White, Scott and Silk, 1994). Many 
distortions of the standard model of recombination have also
been investigated. The delay of recombination due to 
evaporation of primordial black holes has been discussed in
Naselsky (1978) and Naselsky and Polnarev (1987). Avelino et al.
(2000), Battye et al. (2001), Landau et al. (2001) have noted
that possible time variations of the fundamental constants could
also influence the ionization history of the cosmic plasma. Sarkar 
and Cooper (1983), Scott et al. (1991), Ellis et al. (1992), Adams 
et al. (1998), Peebles, Seager and Hu (2000), Doroshkevich and Naselsky 
(2002) discussed distortions of the kinetics of hydrogen recombination 
caused by decays of hypothetical particles. It is worth 
noting that for such models the energy injection delays the 
recombination at $z\simeq 10^3$ and distorts the ionization 
history of the Universe down to the period of galaxies formation 
at $z\simeq 5-10$. 

Recently Naselsky and Novikov (2002) have proposed 
a model of accelerated recombination with non--uniform 
spatial distribution of baryonic matter. In this model 
the recombination proceeds faster within the clumps and slowly 
in the intercloud medium. In such models at redshifts $z\simeq 
10^3$ the mean ionization fraction decreases faster than that 
in the standard model with the same cosmological parameters 
which mimics the acceleration of recombination. Here we show that, 
for suitable parameters of the clouds, such models can improve the 
best--fit of the observed power spectrum of the CMB anisotropy.
 
The non-standard models of hydrogen recombination are 
characterized by a few additional parameters which must be 
included in the list of the best-fit cosmological
parameter determination using the CMB anisotropy and
polarization data. We call these parameters as the ``missing 
parameters'' of the theory and show how important  
these parameters may be for the CMB physics. 

In this paper we compare the observed CMB anisotropy power 
spectrum for the reference model of the standard recombination, 
with models of delayed and accelerated recombination.
For the reference model we accept the cosmological parameters 
from the best-fit of the CBI observational data (Mason et al 2002). 
We show that the models considered in this paper can provide better 
fits for the already observed and the future MAP and PLANCK 
power spectra and can change usual estimates of the cosmological
parameters. Moreover, the measurements of the polarization allow
us to discriminate between some of such complex models of the
Universe. These results can be important for the interpretation
of the MAP and upcoming PLANCK data.

The paper is organized as follows. In section 2 we discuss
the generic models of distorted recombination. In section 3 
we demonstrate distortions of the CMB anisotropy power spectrum 
due to delay or acceleration of recombination. In section 4 we 
introduce the models of accelerated and delayed recombination 
and discuss the redshift variations of the hydrogen ionization 
fraction in these models. In section 5 we compare the observed 
CMB anisotropies and expected polarization power spectra with 
ones for the models under discussion. Section 6 is devoted to 
discussion of late reionization and of the 
polarization power spectrum for the delayed recombination models. 
In Conclusion we make some predictions for the MAP and the 
upcoming PLANCK experiments.

\section{Deviations from the standard model of hydrogen 
recombination}

The deviations from the standard recombination process 
caused by the injection of additional Ly-$\alpha$ and Ly-c
photons at the recombination epoch or by strong small
scale clustering of the baryonic component 
can be suitably described in terms of 
an additional source of Ly-$\alpha$ and Ly-c photons which was 
proposed in Peebles, Seager and Hu (2000). In the generalized 
recombination model the rates of production of resonance, 
$n_{\alpha}$, and ionized, $n_{i}$, photons are described
by functions $\varepsilon_{\alpha}(z)$ and $\varepsilon_{i}(z)$,
defined as follows:
\be
\frac{dn_\alpha}{dt}=\varepsilon_\alpha(z)H(z)\langle n_b(z)\rangle\,,
\label{alp-i}
\ee
 \be
\frac{dn_i}{dt}=\varepsilon_{i}(z)H(z)\langle n_b(z)\rangle\,,
\label{i-alp}
\ee
where $H(z)$ and $\langle n_b(z)\rangle$ are the Hubble parameter 
and the mean baryonic density.
 
\subsection{Delay of recombination}

 Peebles, Seager and Hu (2000) considered the models with 
$\varepsilon_\alpha(z)=const.$ and $\varepsilon_i(z)=const.$  
However, in general case both parameters $\varepsilon_\alpha(z)$ 
and $\varepsilon_i(z)$ could depend on redshift (see, e.g., 
Doroshkevich\,\&\,Naselsky 2002; Naselsky\,\&\,Novikov 2002). 
As was shown in these papers, for the generation of ionized photons 
from the decay of super heavy dark matter (SHDM) particles we get 
\be
\varepsilon_\alpha(z)\simeq\varepsilon_i(z)\propto (1+z)^{-1}\,.
\label{eqd2}
\ee 
For models of particle or strings decays the particle 
number density, $n_x$, is a decreasing function of time defined by 
the life time of the particles, $\tau_x(n_x,t)$, as follows:
\be
\frac{dn_{x}}{dt}+3H(t)n_x=-\frac{n_x}{\tau_x(n_x,t)}
\label{eq2}
\ee
For the simplest models with $\tau_x=const.$ we have
\be
\frac{d[n_x(1+z)^3]}{dt}=\varepsilon_{x}(t)H(t) \langle n_b\rangle\,,
\label{eq4}
\ee
\be
\varepsilon_{x}= - \frac{1}{H(t)\tau_x}\exp\left(-\frac{t-t_u}
{\tau_x}\right)\frac{n_x(t_u)}{\langle n_b(t)\rangle}\,
\nonumber
\ee
where $t_u$ is the age of the Universe. 

Obviously, $\varepsilon_i=\kappa_i\varepsilon_x,\, \varepsilon_\alpha= 
\kappa_\alpha\varepsilon_x$ where factors $\kappa_i$ and $\kappa_\alpha$ 
characterize the efficiency of resonance and ionized photons production. 
The decays of $x$-particles cannot significantly distort the 
thermodynamics of the Universe at high redshifts $z\geq 10^3$ but 
this energy injection changes the kinetic 
of recombination at $z\sim 10^3$ that leads to the observable 
distortions of the power spectra of CMB anisotropy and 
polarization. To evaluate these distortions we firstly have 
to follow the transformation of high energy injected particles 
to UV photons influenced directly on the recombination process. 

The electromagnetic cascades initiated by the ultra--high energy 
cosmic rays (UHECR) are composed of photons, protons, electrons, 
positrons and neutrinos. At high redshifts, the cascades develop 
very rapidly via interaction 
with the CMB photons and pair creation ($\gamma_{UHECR}+\gamma_{CMB}
\rightarrow e^{+} + e^{-}$), proton-photon pair production ($p_{UHECR} 
+ \gamma_{CMB}\rightarrow p^{'} + \gamma{'}+e^{+} + e^{-}$), inverse 
Compton scattering ($e^{-}_{UHECR}+\gamma_{CMB}\rightarrow e^{'} + 
\gamma^{'}$), pair creation ($e^{-}_{UHECR}+\gamma_{CMB}\rightarrow 
e^{'}+ e^{-} + e^{+} + \gamma^{'}$), and, for neutrino, electron-
positron pair creation through the Z-resonance. As was shown by 
Berezinsky et al. (1990) and Protheroe et al. (1995), these 
processes result in the universal normalized spectrum of a cascade 
with a primary energy $E_\gamma$ which can be written as follows:
\be 
N_\gamma(E, E_{\gamma})=F(E, E_{\gamma})
\left\{
\begin{array}{cc}
\sqrt{E\over E_a}&E\leq E_a\cr
1&E_a\leq E\leq E_c\cr
0&E_c\leq E\cr
\end{array} 
\right.      
\label{eq1e}
\ee
\bedm
F(E, E_{\gamma})=\frac{E_\gamma E^{-2}}{2+\ln(E_c/E_a)},\hspace{0.5cm}
\int_0^{E_\gamma}EN_\gamma dE=E_\gamma
\eedm
where $E_c\simeq 4.6\cdot 10^4(1+z)^{-1}$GeV, $E_a=1.8\cdot 10^3
(1+z)^{-1}$GeV. At the period of recombination $z\sim 10^3$ and 
for lower redshifts both energies, $E_a$ and $E_c$, are larger 
than the limit of the electron-positron pair production $E_{e^{+},
e^{-}}=2 m_e = 1$ MeV and the spectrum (\ref{eq1e}) describes both 
the energy distribution at $E\geq E_{e^+,e^-}$ and the injection of 
UV photons with $E\ll E_{e^+,e^-}$. However, the spectrum of these 
UV photons is distorted due to the interaction of photons with the 
hydrogen - helium plasma.  

Let us briefly discuss the relationship between the initial spectra 
of the SHDM decay and the parameters $\varepsilon_{\alpha}$ 
and $\varepsilon_{i}$ of hydrogen excitation and ionization. 
Evidently, the energy density of UV photons is proportional to 
the rate of energy injection, $\dot{Q}(t)$,  
\begin{equation}
N(E)\propto \dot{Q}(t)(E_a/E)^{3/2}= \omega(t)\dot{Q}(t_u)
(E_a/E)^{3/2}\,,
\label{a}
\end{equation}
\[
\dot{Q}(t)=\omega(t)\dot{Q}(t_u)=\int_0^\infty dE~E N(E)\,,
\]
where $\dot{Q}(t_u)$ is the rate of energy injection at $t=t_u$.
For the net of ionizing UV photons with $E\geq I\simeq$ 13.6 eV we have
\begin{equation}
\dot{n}_i\simeq \frac{\omega(t)\dot{Q}(t_u)}{I}\left(\frac{I}
{E_a(z)}\right)^{\frac{1}{2}}
\label{a2}
\end{equation}
Using normalization $\dot{Q}(t_u)H^{-1}(t_u)\simeq \epsilon_{EGRET}$ 
to the EGRET energy density $\epsilon_{EGRET}\simeq 4\cdot10^{-7} 
eV cm^{-3}$ (Berton, Sigl\,\&\,Silk 2001) we can estimate the functions 
$\varepsilon_i$ and $\varepsilon_\alpha$. Namely, for  
\begin{equation}
\omega(t)\propto t^{4-p}\propto (1+z)^{\frac{3}{2}(4-p)}
\label{a3}
\end{equation}
we get
\begin{equation}
\varepsilon_i\simeq \varepsilon_\alpha\simeq \xi (z)
\left(\frac{ \epsilon_{EGRET}}{In_b(z=0)}\right)(1+z)^{\frac{3}{2}(1-p)} 
\propto (1+z)^{2(1-3p/4)}
\label{a4}
\end{equation}
\[
\xi(z)= \left(\frac{I}{E_a(z)}\right)^{\frac{1}{2}}\simeq 
3\cdot10^{-6}\sqrt{1+z}
\]

The power index $p=1$ corresponds to the model involving the release 
of $x$-particles from topological defects, such as ordinary cosmic 
strings, necklaces and magnetic monopoles (see, e.g., Sigl 2001). 
The model with $p=2$ corresponds to the exponential decay of SHDM 
particles described by Eq.~(\ref{eqd2}) with $\tau_x\gg t_U$. The 
model discussed in Peebles, Seager and Hu (2000) with 
$\varepsilon_i=const.$ and $\varepsilon_\alpha(z)=const.$ 
corresponds to $p=4/3$.

For some models such as those of evaporation of primordial black 
holes or decays of SHDM relics with shorter life time, 
$\tau_X=t(z_x)\le t_u$, more complex normalization is required. 
Thus, for $\tau_X\simeq 0.1 t_u$ that corresponds to redshift 
$z_x\sim 6-7$, these decays occur before or just at the epoch 
of galaxy formation and now their number is small. For 
such models and for $z\gg z_x$ we get, instead of Eq.~(\ref{a4}), 
\be
\xi_{mod}(z)=\xi(z)\Theta(z_x)=\xi(z)\exp\left((1+
z_x)^{\frac{3}{2}}\right)\,.
\label{aa4}
\ee

For some models with different $p$ and $\Theta(z_x)$, functions 
$\varepsilon_\alpha(z)$ are plotted in Fig. 1. As one can 
see, for some of the models we should have $\varepsilon_{\alpha}\ge 
10^{-3}-10^{-2}$ at $z<1500$ and distortions of the kinetics of
hydrogen recombination can be quite strong. For other models these 
distortions are small and cannot be observed. 

Other approach to restrictions of admissible energy injection 
is discussed in Sec. 6\,.
 
\subsection{Acceleration of recombination}

The basic idea of accelerated recombination is very transparent 
(Naselsky and Novikov 2002). Let us consider the model in which 
a fraction of baryons is concentrated in high density low 
mass clouds. The hydrogen 
recombination inside the clouds proceeds faster than in the low density 
intercloud medium, and the mean ionization fraction is 
\be
\langle x_e(z)\rangle=x_{e, in}Z_m + x_{e, out}(1-Z_m)
\label{x}
\ee
where $Z_m$ is the mass fraction accumulated by clouds, $x_{e, in}$ 
and $x_{e, out}$ are the ionization fractions within and between clouds. 
This expression clearly shows that in such models the redshift variations 
of $\langle x_e(z)\rangle$ differ from those in the standard models 
with homogeneous spatial distribution of baryons and at the same 
values of cosmological parameters.  

Cosmological models with such clumpy distribution of 
baryons appear when we consider the mixture of adiabatic and 
isocurvature (or isotemperature) perturbations (Dolgov\,\&\,Silk 
1993). In suitable models 
isocurvature perturbations provide the concentration of some 
fraction of baryons within low massive clouds with $M\sim 10^2-10^5 
M_{\odot}$ while adiabatic perturbations dominate on larger scales 
and are responsible for galaxy formation and other effects.  

The ionization history of hydrogen in such models can also 
be described by Eqs.~(\ref{alp-i},\,\ref{i-alp}) with 
$\varepsilon_\alpha\leq 0$ and $\varepsilon_i\geq 0$ (Naselsky 
and Novikov 2002). Of course, this choice does not mean that 
Ly-$\alpha$ photons disappear or are really transformed to Ly-c 
ones. But the suitable choice $\varepsilon_\alpha\leq 0$ allows 
to describe correctly the faster hydrogen recombination within 
the clouds while the choice $\varepsilon_i\geq 0$ allows for the 
excess of Ly-$\alpha$ photons and the delay of hydrogen 
recombination in the low density intercloud medium. 

\subsection{Three scenarios of recombination distortions}

For wide set of models, the functions $\varepsilon_\alpha$ and 
$\varepsilon_i$ can be written as follows:
\be
\varepsilon_{i,\alpha}(z)= \sum_j \exp\left[-\zeta^{m_j}\right]
c^{i,\alpha}_j\zeta^{n_j},\quad \zeta=(1+z)/(1+z_d)\,,
\label{eq5}
\ee
where parameters $c^{i,\alpha}_j, z_d, m_j,$ and $n_j$ 
characterize the $j^{th}$ source of distortion.
It is natural to expect that $\varepsilon_i(z)\geq$ 0 for all redshifts
$z\leq 10^3$. This additional ionization suppresses partly
the CMB anisotropy because of the growth of the optical depth for
the Compton scattering at $z\leq 10^3$. 

Three scenarios can be discussed depending upon the shape of
the function $\varepsilon_\alpha(z)$. Namely, the models with 
$\varepsilon_\alpha(z)\geq$ 0 and $\varepsilon_\alpha(z)\leq$ 0 
describe the delay or acceleration of recombination, respectively. 
More complex models with the acceleration of the recombination at 
$z\geq z_{ad}$ and the delay of recombination at $z\leq z_{ad}$ are 
described by complex functions with $\varepsilon_\alpha(z)
\leq$ 0 at $z\geq z_{ad}$ and $\varepsilon_\alpha(z)\geq$ 0 for 
$z\leq z_{ad}$.

All these scenarios can be based on the realistic physical models.
For example, for the decay of the long lived SHDM particles 
we get
\be
\varepsilon_\alpha\sim\varepsilon_{i}\propto (1+z)^{-1}\,,
\label{shdm}
\ee
which corresponds to the first scenario. The same scenario is 
realized for the decay of topological defects and evaporation 
of primordial black holes discussed above. In all the 
cases, the Ly-c and Ly-$\alpha$ photons are formed via the 
electromagnetic cascades and $\varepsilon_\alpha\sim\varepsilon_i$. 
Third scenario is realized for the clumpy baryonic model discussed 
above. The list of the physical models can be essentially extended.

\section{Distortion of the CMB anisotropy power spectrum}

As was mentioned in Peebles, Seager and Hu (2001), the distortions 
of recombination shift the redshift of the last scattering 
surface, $z_r$, and change its thickness, $\Delta_z$. In turn, these 
variations shift positions of the peaks in the CMB anisotropy 
and polarization power spectrum (Hu and Sugiyama 1995) and change 
their amplitudes. Here we give the rough analytical estimates of 
these influences. More accurate numerical results are presented 
below. 

Both parameters of the last scattering surface, $z_r$ and $\Delta_z$, 
are roughly expressed through the Thomson optical depth, $\tau_T$, 
and its derivations, $\tau^{'}_T= d\tau_T/dz$ : 
\be
\tau_T=\int_0^{z_r} \tau^{'}_T(z) dz,\quad  
\tau^{'}_T(z)={c\sigma_T x_e(z)\langle n_b(z)\rangle\over H(z)(1+z)}
={x_e\Phi(z)\over 1+z}\,.
\label{g1}
\ee
Here $\sigma_T$ is the Thomson cross-section and $c$ is the speed 
of light.

The redshift of the last scattering surface, $z_r$, is defined 
by the position of the maximum of the so--called visibility function, 
$g(\tau_T)=\tau^{'}_T\exp(-\tau_T)$,
\be
\frac{dg(\tau_T)}{dz}|_{z=z_r}=0,
\quad \tau^{''}_T(z_r)=\left(\tau^{'}_T(z_r)\right)^2\,,
\label{g0}
\ee
and its width is
\be
\Delta_z\simeq\left(2\left(\tau^{'}_T(z_r)\right)^2 - 
\frac{\tau^{'''}_T(z_r)}{\tau^{'}_T(z_r)}\right)^{-1/2}
\label{g2}
\end{equation}
(see, e.g., Naselsky \& Polnarev 1987). Taking into account that  
recombination is a fast process and $\Delta_z\simeq 0.1 z_r$, 
we can further consider only redshift variations of 
the ionization fraction, $x_e$, and take the function $\Phi$ in 
(\ref{g1}) at $z=z_r$. This means that the parameters 
$z_*$ and $\Delta_z(z_*)$ can be found from the following equations:  
\be
x^{'}_e(z_*)\simeq x_e^2(z_*)\Phi(z_r)/z_r,\quad 
\Delta_z\sim x_e(z_*)/x^{'}_e(z_*)\approx x_e^{-1}(z_*)z_r/\Phi(z_r)\,.
\label{g4}
\end{equation}
Here $ z_r$ and $z_*$ are the redshifts of the last scattering 
surface in the standard and the distorted models. For small distortions 
of ionization fraction we have 
\[
x_e\approx x_{e,st}+x^{'}_{e,st}z_r(1-z_*/z_r),\quad 
x^{'}_e\approx x^{'}_{e,st}-n_b^{-1}dn_\alpha/dz=
x^{'}_{e,st}-\varepsilon_\alpha(z_r)/z_r\,,
\]
\be
z_*\simeq z_r(1-\varepsilon_\alpha(z_r) \nu),\quad 
\nu=0.5\Phi^{-2}(z_r)x_{e,st}^{-3}\,,
\label{g7}
\ee
\[
\Delta_z(z_*)\sim \Delta_{z,st} [1+0.5\varepsilon_\alpha(z_r) 
\Phi^{-1}(z_r)x_{e,st}^{-2}]\,. 
\]
Here $x_{e,st}$ and $\Delta_{z,st}$ are the ionization fraction 
and the thickness of the last scattering surface for the standard 
model. 

The leading order dependence of the first peak location is 
(Peebles, Seager and Hu 2001)
\be
l_1(\varepsilon_\alpha)\simeq l_{1, st}\sqrt{z_*/z_r}\simeq 
l_{1, st}[1 - 0.5\nu\varepsilon_\alpha(z_r)]\,.
\label{g7}
\end{equation}
For the second peak of the CMB power spectrum the shift of the 
position is practically the same as for the first one, but the 
ratio between their amplitudes is a linear function of ratio 
$z_*/z_r$. For the next few peaks growth of the thickness of the last 
scattering surface, $\Delta_z$, increases the damping and decreases 
the amplitudes of the peaks (White 2001). In contrast, for 
models with accelerated recombination and $\varepsilon_\alpha\leq 0$, 
peaks are shifted to larger $l$ and the damping of perturbations 
decreases together with $\Delta_z$ as compared with the standard 
model. 

\section{Five models of recombination and distortions of the 
ionization history.}

\subsection{Models of recombination.}

To illustrate the possible manifestation of the external sources 
of resonance and ionized photons and non--uniform distribution of 
baryonic matter in the ionization history of the Universe
we compare four models with different ionization history with 
the basic spatially flat $\Lambda$CDM model, $\Omega_K=0$, 
the standard ionization history, $\varepsilon_i=\varepsilon_
\alpha=0$, and 
\be
\Omega_\Lambda=0.7,\quad \Omega_c+\Omega_b=0.3,\quad h=0.7,\quad 
n_s=1,\quad \Omega_bh^2=0.022,\quad \Omega_ch^2=0.125\,.
\label{mparam}
\ee
This baryonic matter density is close to the most probable 
value found from the comparison of the Big-Bang nucleosynthesis 
prediction with the observable abundance of the light chemical 
elements. The basic model includes also the optical depth for the Compton 
scattering due to reionization of the Universe by young galaxies 
and quasars at redshift $z_{rei}$	
\be
\tau_{rei}=63 x_e\left(\frac{1-Y_p}{0.76}\right)
\left(\frac{\Omega_bh^2}{0.02}\right)
\left(\frac{\Omega_ch^2}{0.122}\right)^{-\frac{1}{2}}
\left(\frac{1+z_{rei}}{10^3}\right)^{\frac{3}{2}}\,,
\label{mparam1}
\ee  
where $x_e\simeq 1$ is the fractional ionization at the redshift 
$z_{rei}$ and $Y_p$ is the mass fraction of helium.  For parameters 
(\ref{mparam}), $Y_p=0.24$, $z_{rei}\simeq 14$ and $\tau_{rei}\simeq 
0.1$, this model provides the best--fit for the recent CBI data. We 
will refer to the basic model as to Model 1\,.

Recent observations of the farthest quasars (Fan et al. 2001) give the 
low limit for the epoch of reionization, $z_{rei}\ge 6$, which does 
not contradict the adopted value $z_{rei}\simeq 14$. For $z_{rei}
\simeq 6$, the Thompson optical depth is only $\tau_{rei}\simeq 0.04$. 
We will refer to the model with parameters (\ref{mparam}), $z_{rei}
\simeq 6$ and $\tau_{rei}\simeq 0.04$ as to the Model 1a and will 
discuss it in Section 6. 

To demonstrate the influence of distortions of the ionization 
history, we consider three models with various distortions. 
In Model 2 we change the baryonic fraction and accept $\Omega_b
h^2=0.03$ preserving the total density, $\Omega_b + \Omega_c=0.3$.
The Model 3 is the cloudy baryonic model with $\langle\Omega_b 
\rangle h^2=0.022$ and reionization of hydrogen at redshift
$z_{rei}\simeq 14$ as for the Model 1. We assume also that the 
density contrast between clouds, $\rho_{in}$, and interclouds 
medium, $\rho_{out}$, is $\eta=\rho_{in}/\rho_{out}=11$, 
clouds occupy the volume fraction $f_v=0.1$ and accumulate 
the mass fraction $Z_m=\eta f_v\left[\eta f_v+ (1-f_v)\right
]^{-1}=0.55$. For simplicity, we consider all clouds with the 
same mass in range $10^2 M_{\odot}\le M_{cl}\le 10^5 M_{\odot}$ 
(Naselsky and Novikov, 2002).

The Model 4 differs from the Model 1 by action of external sources 
of ionization with     
\bedm
\varepsilon_\alpha=\alpha\xi^{-\frac{3}{2}}
\exp\left(-\xi^{-\frac{3}{2}}\right),\quad 
\varepsilon_i=\beta\xi^{-\frac{3}{2}}
\exp\left(-\xi^{-\frac{3}{2}}\right)\,,
\label{qm}
\eedm
\[
\xi=(1+z)/(1+z_d)\simeq 10^{-3}(1+z)\,.
\]
The choice $\alpha\simeq 0.3$, $\beta\simeq 0.13$ coincides with the 
model of primordial black hole evaporation at the moment of 
recombination (Naselsky 1978; Naselsky \& Polnarev 1987). In this 
Model we exclude the impact of late reionization which mimics 
roughly the model with $z_{rei}\simeq 6$. For some applications we 
consider also the Model 5 which differ from the Model 4 by the 
baryonic density,  $\Omega_bh^2=0.03$ preserving the total density, 
$\Omega_b + \Omega_c=0.3$. In this respect, the Model 5 is identical 
to the Model 2.

Main parameters of the Models are listed in Table 1.

For all the models the anisotropy and polarization power spectra
are found with modified RECFAST and CMBFAST codes. To characterize 
the differences between the models and to compare them with the 
expected sensitivity of the PLANCK experiment we describe 
the deviations of the anisotropy and polarization power spectrum 
in terms of  the functions
\[
D^a_{i,j}(l)=2\left[C^a_i(l)-C^a_j(l)\right]/\left[C^a_i(l)+
C^a_j(l)\right]\,, 
\]
\be
D^p_{i,j}(l)=2\left[C^p_i(l)-C^p_j(l)\right]/\left[C^p_i(l)+
C^p_j(l)\right]\,,
\label{eqd}
\ee
where $i, j$ relate to the Models and $C^a$ and $C^p$ are the 
power spectrum of  anisotropy and $E$ component of polarization, 
respectively. The differences between the ionization fractions in the 
models are characterized by the ratios
\be
\Delta x_{i,j}=2\frac{x_i - x_j}{x_i +  x_j}\,.
\label{eqd1}
\ee

\subsection{Distortions of the ionization history}
 
The redshift variations of the relative difference between the 
ionization fractions in the basic Model 1 and three other models, 
$\Delta x_{1,j},\, j=2, 3, 4$ are plotted in Fig. 2 together with 
the redshift variations of the ionization fraction in the basic 
Model 1. 

For the Model 2, this difference does not exceed $20\%$ at $z\sim 
10^3$ and increases up to $30\%$ at lower redshifts which demonstrates 
a relatively weak influence of the variations of baryonic fraction. 
For the Model 3 with accelerated recombination, the relative 
difference $\Delta x_{1,3}< 0$ at redshifts $z\ge 700$ and the 
ionization fraction decreases with time faster than in the basic 
model. For $z\le 700$ the ionization fraction is larger than that 
in the basic model, $\Delta x_{1,3}> 0$, but the excess is less 
than $10\%$ and the distortions of the CMB anisotropy and polarization 
power spectra remain small due to the small optical depth.

The conspicuous delay of recombination and distortions of the 
ionization fraction appear for the Model 4. They achieve 
$\sim 20-30\%$ already at $z\simeq 10^3$ and are maximal at 
$z\simeq 500$, which provides the essential growth of the optical 
depth for the Thomson scattering and suppresses the amplitude of the
CMB anisotropy.

These results indicate that we can expect noticeable variations of 
the CMB anisotropy and polarization power spectra for the Model 3 
and the Model 4 while for the Model 2 they can be quite moderate. 

\section{Anisotropy and polarization power spectra as a test of 
the ionization history of the Universe.}

To obtain the power spectra of the CMB anisotropy and
polarization in our models we have to use the modification
of the CMBFAST code (Seljak and Zaldarriaga, 1996) taking
into account more complicated ionization history of the
Universe discussed above. For the reference Model 1 and 2, 
we use the standard CMBFAST code for the optical depth 
$\tau_r$ and the ionization fraction $x_e=1$ achieved at the 
redshift of reionization $z_{rei}$. 

In Fig.3 we plot the CMB power spectrum for the models 1, 3 and 
4 in comparison with the observational data from the BOOMERANG, 
MAXIMA-1, and CBI (Mason et al, 2002) experiments at the multipole 
range $l\leq$ 2 000 where the possible influence of the 
Sunyaev-Zel'dovich clusters is not yet crucial.

As one can see from Fig. 3, for all models the power spectra 
are very similar to each other and the Model 1, 3 and 4 are 
consistent with the available measurements of $C(l)$. This 
means that the more complicated ionization history of hydrogen 
violates the standard fitting procedure. In particular, the comparison 
of results listed in Table 2 for the Model 1, 2, and 5 shows 
that the determination of baryonic density from the CMB 
measurements is ambiguous. This fact is important for the 
interpretation of the MAP and PLANCK measurements.

To characterize the differences between the Models and to
compare them with the expected sensitivity of the PLANCK
experiment we plot in Fig. 4 the functions $ D_{13}(l)$ and 
$ D_{14}(l)$ defined by (\ref{eqd}) for the multipole range 
$2\le l \le$ 2000. These functions can be directly compared 
with the expected error bars of $C(l)$ for the PLANCK 
mission.

As is clear from Fig. 4 for the multipole range $500\leq l\leq 
2000$, $0.05\leq D_{13}(l)\leq 0.2$ and $0.05\leq D_{14}(l)\leq 
0.3$, and there is a quite regular tilt at high multipole range. 
Obviously, the tilt is related to the damping of the $C(l)$ power 
spectrum in the basic model because of the late reionization with
$\tau_{rei}=0.1$. In order to clarify the manifestations of the 
more complex ionization regimes in the Model 3 and 4 we have to 
compare the peak--to--peak variations of amplitudes, $ D_{13}(l)$ 
and $ D_{14}(l)$, with the expected errors of measurements for 
the PLANCK experiment. If the systematic effects will be 
successfully removed, the error should be close to 
\be
\frac{\Delta C(l)}{C(l}\simeq \frac{1}{\sqrt{f_{sky}(l+
\frac{1}{2})}}\left[1+w^{-1}C^{-1}(l)W^{-2}_l\right]
\label{c}
\ee
\[
w=(\sigma_{p}\theta_{FWHM})^{-2},\quad W_l\simeq \exp\left[-
\frac{l(l+1)}{2l^2_s}\right],\quad f_{sky}\simeq 0.65\,. 
\]
Here $f_{sky}$ is the sky coverage during the first 
year of observations, $\sigma_{p}$ is the sensitivity per resolution 
element $\theta_{FWHM}\times\theta_{FWHM}$ and $l_s=\sqrt{8\ln2}
\theta^{-1}_{FWHM}$.

For example, for the HFI 353 GHz channel the $\theta_{\rm FWHM}$ 
is 5 arcmin and the error at $l\le 2000$ are defined by the first 
term in Eq~(\ref{c}) . As one can see from Fig. 4, at $l\sim 1000-1500$ 
the peak--to--peak amplitudes are $\sim 10-20\%$, while $\Delta 
C(l)/C(l)\simeq 5-7\%$. This means that such peculiarities of the 
anisotropy power spectrum can be detected by the PLANCK mission, 
if the systematic effects would be less than $\Delta C(l)/C(l)
\simeq [f_{sky}(l+1/2)]^{-1/2}$.

The polarization power spectra are even more sensitive 
to distortions of the ionization history of the Universe. 
Recently the DASI experiment (Kovac et al. 2002) has detected the CMB 
polarization without significant manifestation of polarized 
foregrounds. It indicates that the launched MAP and the PLANCK 
missions will be able to measure the polarization of the CMB with 
unprecendent accuracy. 

The polarization power spectrum $\Delta T^2_p(l)=l(l+1)C_p(l)/2\pi$ 
are plotted in Fig.5 for the Model 1, 3 and 4 and their relative 
variations, $D^p_{1,j}(l)$, with respect to the basic model are 
plotted in Fig. 6 . As one can see the differences between spectra 
are quite small especially at $l\gg 200$. But the bump at $l\sim 5-10$ 
which is caused by the late reionization in the basic model can be 
essentially suppressed for the Model 4 wherein the required optical 
depth is gained at redshifts $z\sim 100- 500$. 

For the PLANCK mission, even if all the systematic and foreground 
contaminations are removed from the polarization map (see Franco, 
Fosalba and Tauber 2002), one of the most important sources of 
uncertainty at $l\ll 200$ is the cosmic variance. For $f_{sky}
\simeq 0.65$ and $l\simeq 2000$ the cosmic variance limit is close 
to $\Delta C(l)/C(l)\simeq 3\%$ for $1\sigma$ confidence 
level and all peculiarities of the polarization power spectrum should 
be observable. As it follows from Fig. 6, all polarization features 
which corresponds to the Model 1, 3 and 4 can be observed by the 
PLANCK mission, and, for the low multipole part $l\le 200$, can be 
probably observed already by the MAP mission. This means that future 
polarization experiments can be crucial for investigation of the 
ionization history of the cosmic plasma as near the redshift 
of the last scattering surface as at the redshift of reionization. 

\section {Low multipole peculiarities of the polarization power 
spectrum in models with delayed recombination }

The signature of late reionization is the bump at small multipoles, 
$l\le 10$, in the polarization power spectrum. Its amplitude and 
shape depend upon the ionization history characterized by $z_{rei}$
and $\tau_{rei}$. This bump is clearly seen in Fig. 5 for the basic 
model with $z_{rei}=14$ and $\tau_{rei}=0.1$ but it is strongly 
suppressed for the later reionization such as, for example, 
for $z_{rei}\simeq 6$ and $\tau_{rei}=0.04$ widely discussed in 
connection with the recent observations of the farthest quasars 
(Fan et al. 2001). For the non--standard ionization history discussed 
in this paper, the shape of the polarization power spectrum is more 
complicated but its observations allow to restore some features of 
the ionization hystory.  

As is seen from Fig. 6, for the Model 3 with accelerated 
recombination this bump is just the same as in the basic Model 1. 
However, in the Model 4 with delayed recombination and without any 
sudden reionization at low redshifts significant variations of 
$D^p_{1,j}(l)$ are seen in the Fig. 6 at $l\leq 500$. They can 
be observed by the PLANCK experiment even for relatively small 
$\varepsilon_i(z)$, which would clarify the ionization history for 
such models. 

To illustrate these abilities, we plot in Fig. 7 the relative 
distortions of the polarization power spectrum at $l\leq 1000$ 
for several models with delayed recombination and different 
$\varepsilon_i(z)$. First of all, to illustrate the influence 
of the optical depth, $\tau_{rei}$, we compare the Model 1
with the Model 1a with the same parameters and $z_{rei}\simeq 6$
and $\tau_{rei}=0.04$. Large values $D^p_{1,j}(l)$ for these 
models at $l\leq 20$ allows to easily discriminate between these 
models. 

In the same Fig. 7 we compare also the Model 1a with several models 
of delayed recombination with the same cosmological parameters 
(\ref{mparam}) and with
\be
\varepsilon_i(z)=\varepsilon_i(z_r)\left(\frac{1+z}{1+z_r}
\right)^{2(1-\frac{3}{4}p)}
\label{z}
\ee
where $z_r\simeq 10^3$ is the redshift of the last scattering 
surface. We see that for the models with $p=1,\,\&\,\varepsilon_i
(z_r)= 10^{-2}$ and with $p=2,\,\&\,\varepsilon_i(z_r)= 10^{-3}$ 
the relative distortions, $D^p_{1,j}(l)$, are quite similar and 
can be observed at $l\leq$ 100\,. However, for the model with 
$p=1,\,\&\,\varepsilon_i(z_r)= 10^{-3}$ distortions become small 
and are under observational limits. 
These comparisons demonstrate the abilities of feature polarization 
measurements with respect to restrictions of the ionization history 
of the Universe at $z\leq 10^3$. 

In wide class of such models the injection of ionized photons at 
$z\leq z_r$ increases the ionization fraction and the Thompson 
optical depth with respect to the standard models. In such case, 
the already available limitations of the depth $\tau(z)\leq 
\tau_{thr}\approx 0.1$ restrict the admissible $\varepsilon_i(z)$ 
even at high redshifts where estimates (\ref{a4}) are not 
effective. To estimate the optical depth for such models we 
have to trace numerically the ionization fraction at all redshifts. 
However, even rough analytical estimates are informative and allow 
to reject some models. 

For such rough estimates we use the equation of ionization balance
\be 
\alpha_c\langle n_b\rangle x^2_e\simeq \varepsilon_i(z) H(z)
\label{z1}
\ee
where $\alpha_c\propto T_m^{-0.6}$ is the coefficient of hydrogen  
recombination and $T_m$ is the temperature of the matter. As is well 
known, $T_m\propto 1+z$ for $z\ge z_{eq}\simeq 10^2$ and $T_m\propto 
(1+z)^2$ for lower redshifts and, so, $\alpha_c\propto T_m^{-0.6}\propto 
(1+z)^{-2\omega}$, $\omega\approx 0.3$ for $z\ge z_{eq}$ and 
$\omega\approx 0.6$ for $z\le z_{eq}$\,. 
Thus, for $\varepsilon_i(z)$ given by (\ref{z}) we get for the 
ionization fraction and for the optical depth 
\[
x_e(z)\propto \varepsilon^{\frac{1}{2}}_{i}(z_r)(1+z)^{\mu},\quad 
\mu=-0.75 + \omega + (1-3p/4)\,,  
\]
\be
\tau_r\simeq 36~\nu^{-1} \varepsilon^{1/2}_i(z_r)
\left(\frac{1+z_\tau}{1+z_r}\right)^{\nu}
\left(\frac{1-Y_p}{0.76}\right)\left(\frac{\Omega_b h^2}{0.02}
\right)^{1/2}\left(\frac{\Omega_c h^2}{0.125}\right)^{-1/4}
\label{z2}
\ee
where $Y_p$ is the mass fraction of helium, $\nu=1.5+\mu=0.75 + 
\omega + (1-3p/4)$, and $z_\tau$ is the maximal redshift where 
the ionization balance is described by (\ref{z1}). This relation 
is directly applied for $z_\tau\leq z_{eq}$, $\omega=0.6$ and can 
be easily extended for the case $z_\tau\geq z_{eq}$. 

The observations of farthest quasars show that the reionization 
occurs at $z_{rei}\geq 6$ when $\tau\approx 0.04$ and, so, 
for cosmological parameters (\ref{mparam}), $Y_p=0.24$ and 
$\tau_{thr}\simeq 0.1$ we get 
\be
\varepsilon_i(z_r)\le 10^{-6}\nu^2\left(\frac{1+z_\tau}
{1+z_r}\right)^{-2\nu}\,.
\label{z3}
\ee
As compared with the normalization (\ref{a4}), this expression  
more effectively restricts the possible energy injection at high 
redshifts.

\section{Discussion and conclusion}

The already achieved precision of measurements of the CMB anisotropy 
and especially expected one for the MAP and PLANCK missions allows 
to discriminate wide set of cosmological models and to reveal 
noticeable distortions of the standard model of hydrogen recombination 
process at redshifts $z\simeq 10^3$ and of the ionization history. 
Many physical models can be considered as a basis for such discussions 
and can be tested and restricted via such measurements.

In this paper we compare available observations of the CMB anisotropy
with those expected in several models with different kinetics of
recombination. The differences can be caused by possible external
sources of resonance and ionized radiation, possible small
scale clustering of baryonic component and other factors. 
We show that the models with more complicate ionization history 
provide better description of the available observational data. This  
means also that the accepted estimates of cosmological parameters 
can be changed when such models are incorporated in the 
fitting procedure. It is especially crucial for the 
estimates of baryonic density, $\Omega_b h^2$, which is closely related 
to the Big-Bang nucleosynthesis predictions.   

In this paper we do not consider the impact of Sunyaev--Zel'dovich 
(SZ) effect and the weak lensing by the large scale structure 
elements. However, the SZ effect is frequency dependent that allows 
to reduce its contribution for multifrequency measurements. As was 
shown in Schmalzing, Takada and Futamase (2000), 
the influence of lensing effect can really distort the measured 
characteristics of the CMB only at larger $l\geq 10^3$ while the 
discussed deviations of ionizing history distort these characteristics 
at all $l$. This means that the measurements of the CMB spectra 
in wide range of $l$ allows to discriminate between these 
distortions. However, this problem requires more detailed discussions 
for various models of both lensing and recombination. 

Near the first Doppler peak, the cosmic variance effect restricts an accuracy 
of estimates of $C^a(l)$ as $\Delta C^a(l)/C^a(l)\sim 10\%$ for all 
experiments with the sky coverage $f_{sky}\sim 65\%$. For the recent 
balloon-borne and ground--based experiments the achieved accuracy in this  
range of $l$ is even worse than this limit. With such precision, 
we cannot discriminate the basic model and models with delayed 
recombination and another $\Omega_b h^2$. To illustrate this statement 
we compare quantitatively the observational data with the power 
spectra for Model 1, 2\,\&\,3 parameters which are given in 
Sec. 4.1\,. Instead of the Model 4 we here present the results for 
Model 5 which differs from Model 4 by the values of parameters 
$\Omega_bh^2=0.03$ and $\Omega_m=0.256$ which are the same as in 
the Model 2. 

The quality of models in comparison with the recent BOOMERANG, 
MAXIMA-1, CBI and VSA observational data can be characterized
quantitatively by the $\chi^2$-parameter listed in Table 2 
for all models. Here the reference Model 1 provides the best--
fit for the CBI data for the standard ionization history while
 $\chi^2_2\geq \chi^2_1$ for the Model 2 with larger $\Omega_bh^2$\,. 
However, the Model 3 demonstrates an excellent agreement with 
the CBI data, $\chi^2_3=0.5\chi^2_1$, and $\chi^2_3\leq\chi^2_1$ 
for other observations. For the Model 5 with $\Omega_bh^2=0.03$ 
we get also $\chi^2_{5}\leq \chi^2_1$ for the CBI data and
only for the BOOMERANG data $\chi^2_{5}\geq \chi^2_1$.

Obviously,  such extension of the cosmological models increases
the number of parameters using to fit observed power spectra of the
anisotropy and, in perspective, of the polarization. We show that
the influence of these "missing" parameters, namely $\varepsilon_{i}(z)$
and $\varepsilon_{\alpha}(z)$ , can improve the fits used and, 
in particular, to obtain better agreement between
the observed and expected positions and amplitudes of peaks for higher
multipoles. At the same time, including of the "missing" parameters
changes the measured values of cosmological parameters usually
discussed.

We show also that the expected sensitivity of the MAP and the PLANK 
missions in respect of the measurements of the polarization will 
allow to discriminate between main families of such models and, 
in particular, between models with small and large optical depth 
at redshifts $10^3\geq z\geq$ 20-50. We would like to note that 
the realistic values of the cosmological parameters can not be obtained 
from the CMB data without the PLANCK observations of the polarization.

\section{Acknowledgments}
This paper was supported by Denmark's
Grundforskningsfond through its support for an establishment of
Theoretical Astrophysics Center and, in part, by the Danish
Natural Science Research Council through grant No. 9701841, and
grant RFBR 17625. We should like to thank the anonymous referee 
for useful comments.

\clearpage

\onecolumn

\clearpage

\begin{figure}
\plotone{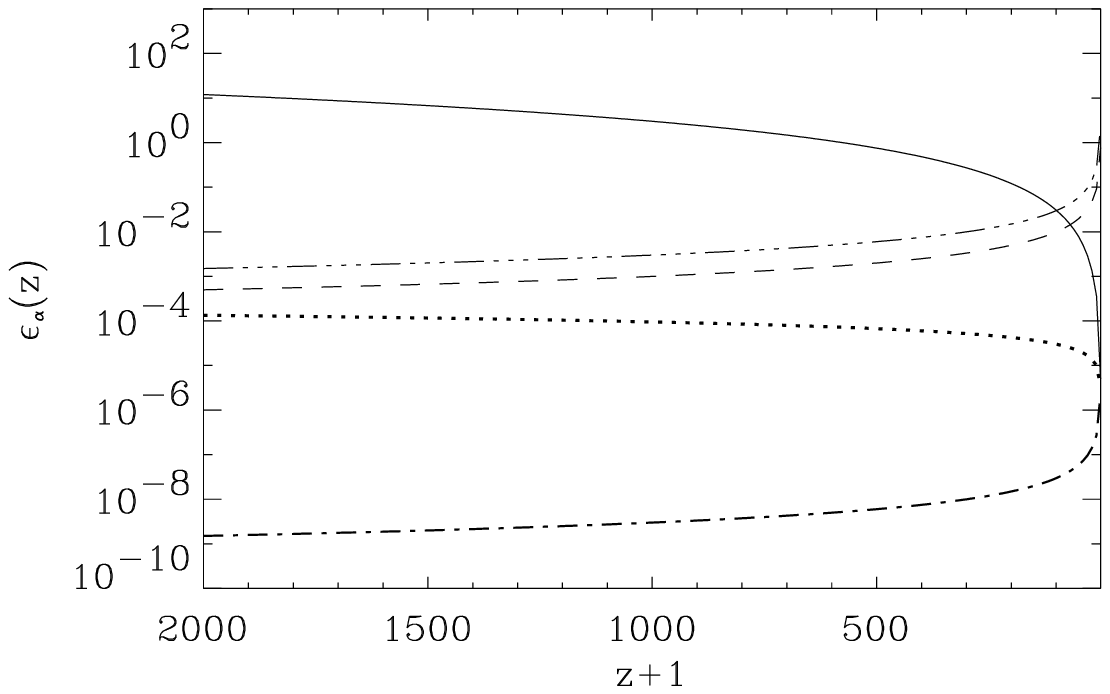}
\vspace{0.2cm}
\caption{Redshift variations of $\varepsilon_{\alpha}$ for models 
(\ref{a4}, \ref{aa4}) for $p=0, ~\Theta(z_x)=1$ (solid line), 
$p=2, ~\Theta(z_x)=1$ (dot-dashed line), $p=2, ~\Theta(z)=1$ 
(dot --dashed line), $p=2, ~\Theta(z)=10^6$ (dot-dot-dot dashed line) 
and $p=2, ~\Theta(z)=3\cdot 10^5$ (dashed line).
}
\label{fig1}
\end{figure}

\begin{figure}
\plotone{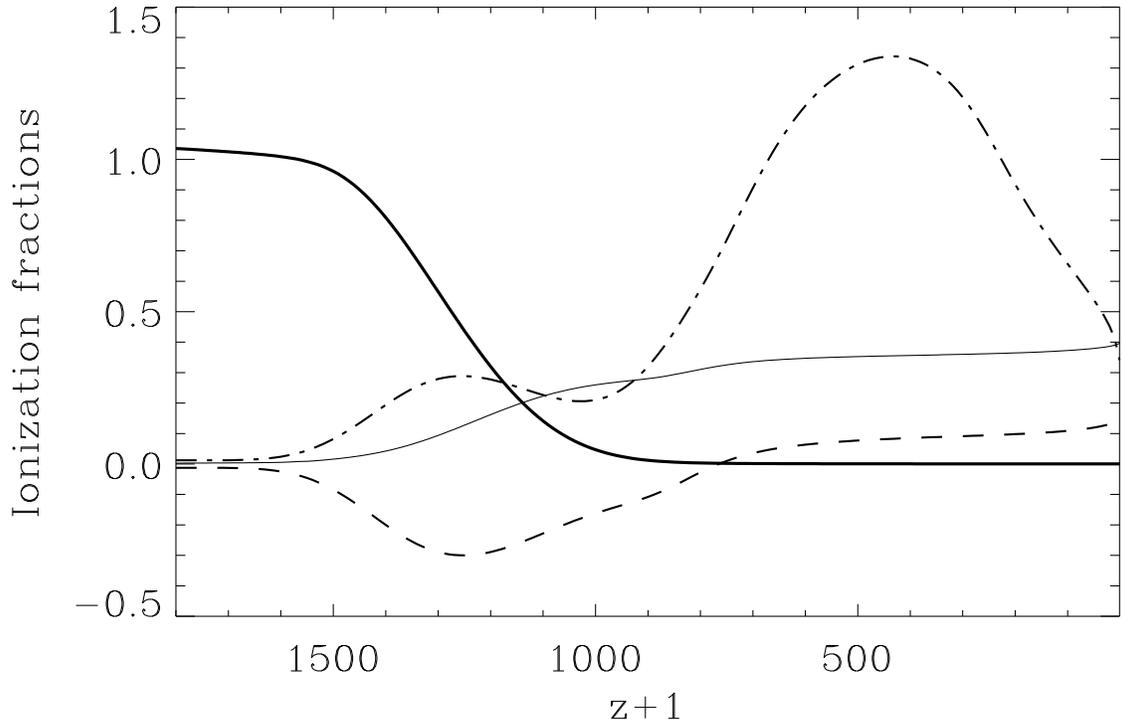}
\vspace{0.05cm}
\caption{Redshift variations of relative differences of the 
ionization fractions between basic Model 1 and other Models, 
$\Delta x_{1,j}$, for $j=2, 3\,\&\,4$ (thin solid, dashed and 
dot-dashed lines, respectively). Thick solid line shows the 
variations of the ionization fraction for the basic Model 1.
}
\label{fig2}
\end{figure}
\clearpage

\begin{figure}
\plotone{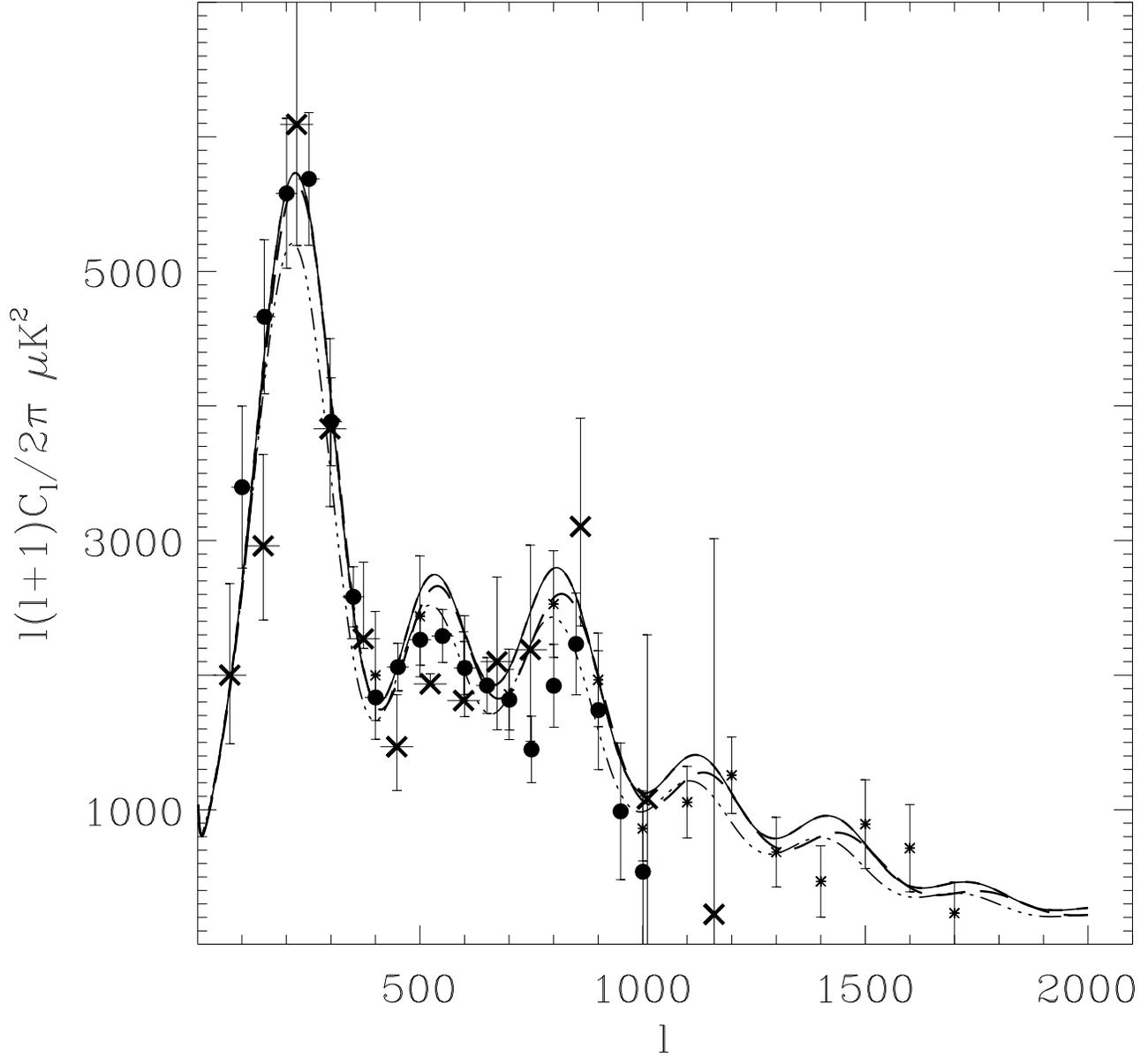}
\vspace{0.05cm}
\caption{CMB anisotropy power spectrum for the Model 1 
(solid line), the Model 3 (long dashed line) and the Model 4 
(dot-dashed line) in comparison with the BOOMERANG (points), 
MAXIMA -1 (large X) and CMI M1 (small x) data.
 }
\label{fig3}
\end{figure}
\clearpage

\begin{figure}
\plotone{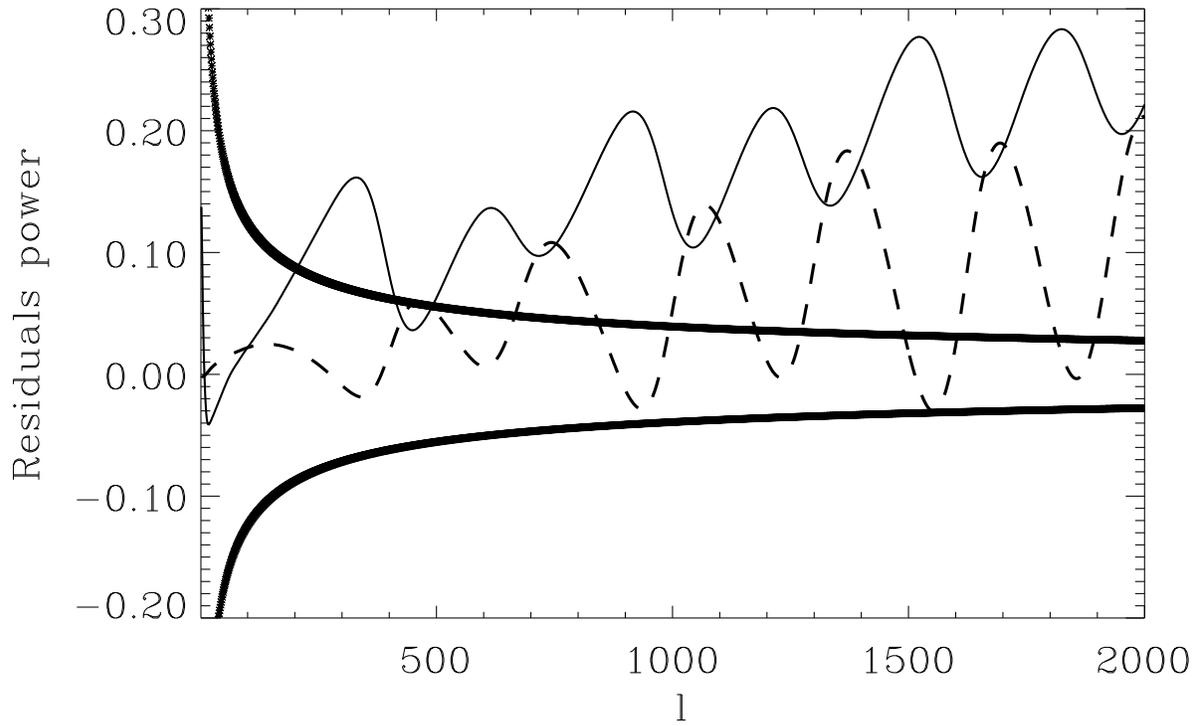}
\vspace{0.05cm}
\caption{Relative differences of CMB polarization power spectra,
$D^a_{1,3}(l)$ (solid line) and $D^a_{1,4}(l)$ (dashed line).  
Thick solid line show expected errors for the PLANCK mission 
(without systematics errors). 
 }
\label{fig4}
\end{figure}
\clearpage

\begin{figure}
\plotone{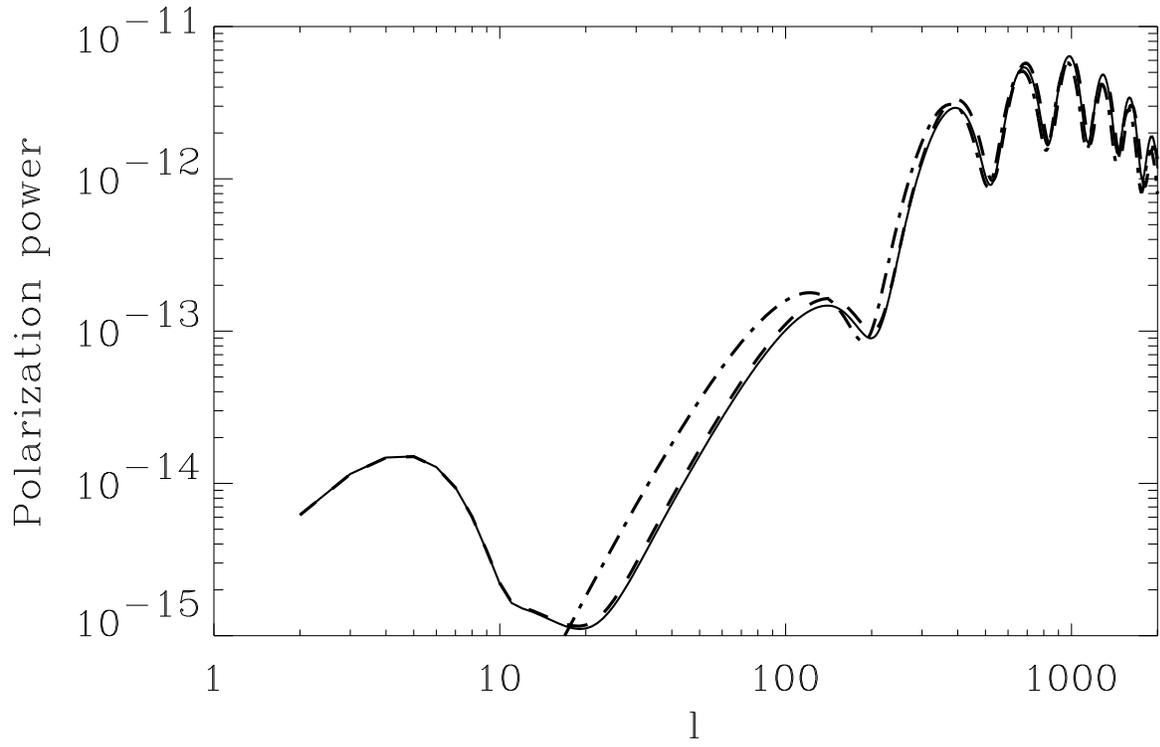}
\vspace{0.5cm}
\caption{CMB polarization power spectra for the Model 1 (solid 
line), the Model 3 (dashed line), and the Model 4 (dot-dashed line).
}
\label{fig5}
\end{figure}
\clearpage

\begin{figure}
\plotone{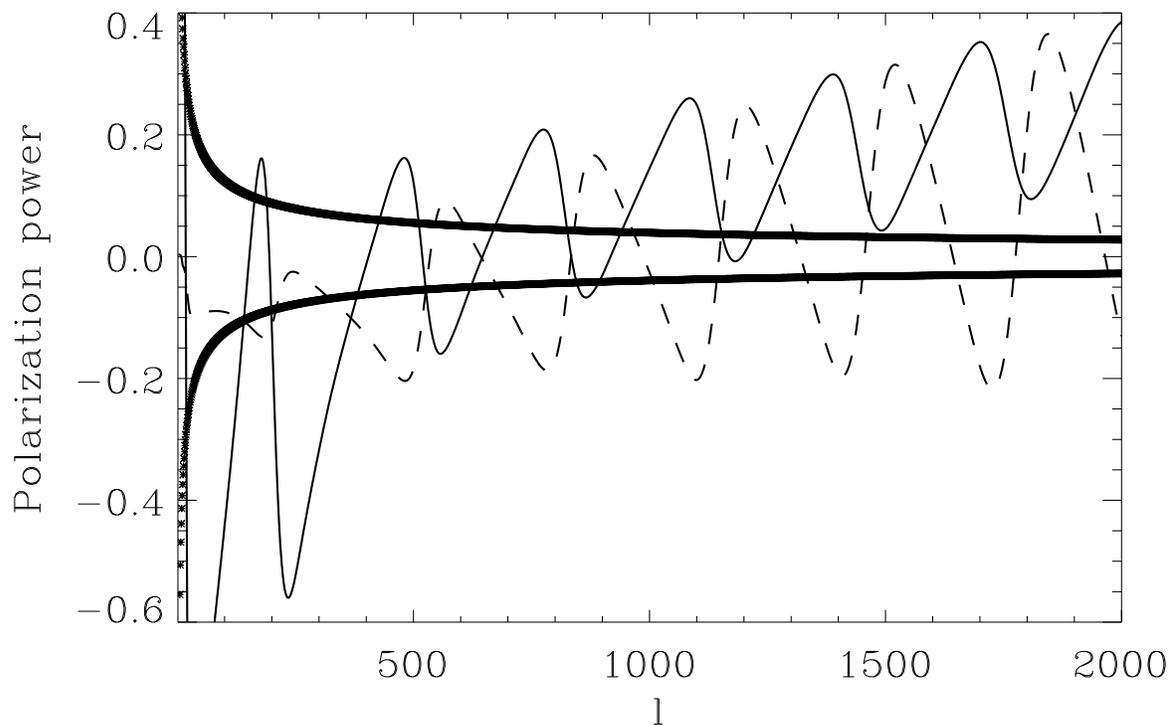}
\vspace{0.5cm}
\caption{Relative differences of the CMB polarization power 
spectrum, $D^p_{1,j}$, between the basic Model 1 and the Model 3 
(dashed line) and the Model 4 (solid line). Thick solid line
show expected errors for the PLANCK mission (without systematics 
errors).
}
\label{fig6}
\end{figure}
\clearpage

\begin{figure}
\plotone{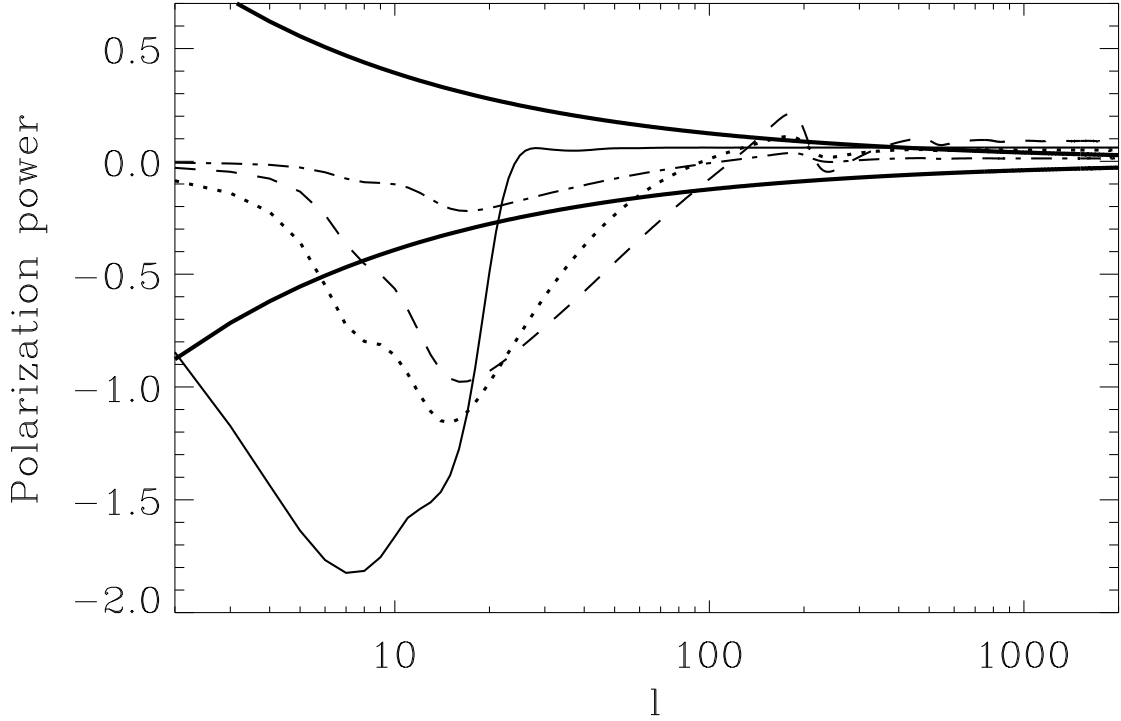}
\vspace{0.5cm}
\caption
{Relative differences $D^p_{1,k}$ of the CMB polarization power 
spectra for low multipoles  for four different models of ionization.
The difference between the basic Model 1 with $\tau_{rei}=0.1$ and 
Model 1a with the same parameters and $\tau_{rei}=0.04$ is plotted 
by solid line; the differences between the Model 1a and models 
(\ref{z}) with $p=1,~ \varepsilon(z_r)=10^{-2}$ and $\varepsilon(z_r)
=10^{-3}$ are plotted by dashed and dot dashed lines; the difference 
between the Model 1a and model (\ref{z}) with $p=2, ~\varepsilon(z_r)
=10^{-3}$ is plotted by dot line. Thick solid lines show expected 
errors for the PLANCK mission.
}
\label{fig7}
\end{figure}
\clearpage

\begin{table}
\caption{Main parameters of the Models}
\label{tbl2}
\begin{tabular}{lrr crl} 
   &   &$\Omega_c$&$\Omega_bh^2$&$z_{rei}$&$\tau_{rei}$\cr
\hline
   1 &basic     &0.256&0.022 &14&0.1\cr
   1a&basic     &0.256&0.022 &6&0.04\cr
   2 &basic     &0.240&0.03  &14&0.1\cr
   3 &accelerate&0.256&0.022 &14&0.1\cr
   4 &delayed   &0.256&0.022 & 0&0.0\cr
   5 &delayed   &0.240&0.03  & 0&0.0\cr
\hline
\end{tabular}
\end{table}

\clearpage

\begin{table}
\caption{$\chi^2$ for five observed anisotropy power spectra 
and for the Models 1, 2, 3,\,\&\,5}
\label{tbl1}
\begin{tabular}{lrr ccr} 
&BOOM &MAX&CBIM1&CBIM2&VSA\cr
\hline
   1 &11.2&14.1 &3.20 &7.15 & 7.1\cr
   2 &34.6&17.7 &2.75 &7.51 &13.0\cr
   3 & 8.7&11.3 &1.60 &5.62 & 6.6\cr
   5 &27.7&15.7 &1.87 &5.15 &10.0\cr
\hline
\end{tabular}
\end{table}


\begin{thebibliography}{}

\bibitem[ad]{ad02} Adams J.A., Sarkar S. \& Sciama D.W., 1998 MNRAS, 301, 210\\
\bibitem[am]{am00} Avelino P., Martins C., Rocha G., \& Viana P., 2000,
Phys.Rev.D, 62,123508\\
\bibitem[bc]{bc01} Battye R., Crittenden R., \& Weller J., 2001, Phys.Rev.D,
63, 043505\\
\bibitem[ben]{ben02} Benoit, A. et al., 2002, Astroparticle Physics, 
17, 101 \\
\bibitem[bbd]{bez}Berezinsky, V. S., Bulanov, S. V., Dogel, V. A., Ginzburg, V. L. and V. S. Ptuskin,
\bibitem[db]{db00} de Bernardis P. et al., 2000, Nature, 404, 955\\
Astrophysics of Cosmic Rays, (North Holland, Amsterdam,1990)\\
\bibitem[bss]{bss01} Berton~G., Sigl~G., Silk~J., 2001, MNRAS, 326, 799
\bibitem[ds]{ds93} Dolgov~A., and  Silk~J., 1993, Phys.Rev. D, 47, 4244
\bibitem[dn]{dn02} Doroshkevich A. \& Naselsky P.D., 2002,  Phys.Rev.D 65,
13517\\
\bibitem[eg]{eg92} Ellis J., Gelmini G., Lopez J., Nanopoulos D.,
\& Sarkar S., 1992, Nucl.Phys.B., 373 , 399\\
\bibitem[fa]{fa02} Fan~X. et al., 2001, astro-ph/0111184, ApJ., in press\\
\bibitem[fr]{ff02} Franco~G., Fosalba~P. and Tauber~J., 2002, astro-ph/0210109
\bibitem[ha]{ha02} Halverson N.W., 2002, ApJ, 568, 38 \\
\bibitem[hh]{hh00} Hanany S., et al, 2000, ApJ, 545, L5\\
\bibitem[jw]{jw85} Jones B.J.T. \& Wyse R., 1985, A\&A, 149, 144\\
\bibitem[kov]{kov02} Kovac J., Leitch E. M., Pryke C., Carlstrom J. E.,
   Halverson N.W., Holzapfel W. L., astro-ph/0209478
\bibitem[lh]{lh01} Landau S., Harari D., \& Zaldarriaga M., 2001, Phys.Rev.D.,
63, 3505 \\
\bibitem[ma]{ma02} Mason B.S. et al., 2002, astro-ph/0205384\\
\bibitem[na]{na78} Naselsky P.D., 1978, Sov. Astron. Lett., 344, 4\\
\bibitem[np]{np87} Naselsky P.D. \& Polnarev A.G., 1987,  Sov. Astron. Lett.,
13, 67 \\
\bibitem[nn]{nn02} Naselsky P.D. \& Novikov I.D., 2002, MNRAS 334, 137 \\
\bibitem[pp]{pp68} Peebles P., 1968, ApJ, 153, 1\\
\bibitem[psh]{psh00} Peebles P., Seager S., \& Hu W., 2000, ApJ, 539, L1 \\
\bibitem[ ]{ } Protheroe, R.J., Stanev, T. , and V.S.Berezinsky, 1995, 
astro-ph/9409004\\
\bibitem[sc]{sc83} Sarkar S., \& Cooper A., 1983,  Phys. Lett. B, 148, 347\\
\bibitem[srs]{srs91} Scott D., Rees M.J. \& Sciama D.W., 1991, A\&A, 250, 
295\\
\bibitem[stm]{stm00} Schmalzing J., Takada M. \& Futamase T., 2000, ApJ, 
544, 83\\
\bibitem[ss]{ss99} Seager S., Sasselov D. \& Scott D., 1999, ApJ, 523, L1\\
\bibitem[sel]{sel96} Seljak U. \& Zaldarriaga M., 1996, ApJ, 469, 437\\
\bibitem[za]{za82} Zabotin N.A. \& Naselsky P.D., 1982, Sov. Astron., 26, 
272\\
\bibitem[zel]{zel68} Zel'dovich Ya.B., Kurt V. \& Sunyaev R.A.,1968,
Zh.Eksp.Theor.Phys., 55, 278\\
\bibitem[wa]{wa02} Watson R.A. et al, 2002, astro-ph/0205378.\\
\bibitem[wh]{wh01} White M., 2001, ApJ, 555, 88
\bibitem[wss]{wss94} White M., Scott D., Silk J., 1994, ARA\&A, 32, 319
\end{thebibliography}
\end{document}